\documentclass[11pt]{article}
\usepackage{hyperref}
\usepackage{cite}
\usepackage{footnote}
\makesavenoteenv{minipage}
\renewcommand{\title}[1]{%
    \bigskip%
    \begin{center}%
    \Large\bf #1%
    \end{center}%
    \vskip .2in}

\renewcommand{\author}[1]{%
    {\begin{center}
    #1
    \end{center}}}
\newcommand{\address}[1]{\vspace{-1.7em}\vspace{0pt}
    {\begin{center}
    \it #1 
    \end{center}}}

\usepackage{amsmath,amsfonts}
\usepackage[dvips]{graphics}
\usepackage[dvips]{graphicx}
\usepackage{subfigure}
\usepackage{epsfig}
\usepackage{amsmath}
\usepackage{amsfonts}
\RequirePackage{color}

\begin{document} 
	\title{Quantum corrected thermodynamics and $P-V$ criticality of self-gravitating Skyrmion black holes} 
	\author
{Yawar H. Khan$\,^{\rm a,b}$,
Prince A. Ganai$\,^{\rm a,c}$,
Sudhaker Upadhyay $\,^{\rm e,f,g, d}$}


\address{ $^{\rm a}${Department of Physics, National Institute of Technology, Srinagar, Kashmir-190006, India }}
\address{ $^{\rm e}${Department of Physics, K. L. S. College, Nawada (a Constituent Unit of Magadh University, Bodh-Gaya)  Bihar-805110,  India }}
\address{ $^{\rm f}${Visiting Associate, Inter-University Centre for Astronomy and Astrophysics (IUCAA) Pune, Maharashtra-411007 }}
\address{ $^{\rm g}${School of Physics, Damghan University, P. O. Box 3671641167, Damghan, Iran }}
\address{$^{\rm b}$\tt iyhkphy@gmail.com}
\address{$^{\rm c}$\tt princeganai@nitsri.com}
\address{$^{\rm d}$\tt sudhakerupadhyay@gmail.com }

	\begin{abstract}
		 In this paper, we study the quantum corrected thermodynamics of a class of black holes generated by self-gravitating Skyrmion models. One such black hole solution is Einstein-Skrymion black hole. We first compute the ADM mass of Einstein-Skyrmion black hole using on-shell Hamiltonian formalism present already in the literature. We then consider non-extended phase space thermodynamics and derive expressions for various thermodynamic quantities like the Hawking temperature, entropy, pressure, Gibbs free energy, and heat capacity. Next, we study the effect of quantum corrections on the thermodynamics of the Einstein-Skyrmion black hole. We observe that apart from leading to stability, the quantum correction induce an AdS (anti-de sitter) to dS (de sitter) phase transition in  Einstein-Skrymion black hole. Treating cosmological constant as the pressure, we determine the $P-V$ criticality of the Einstein-Skrymion black hole and observe that the $P-V$ criticality depends on model parameters $\lambda$ and $K$. The study of $P-V$ criticality done here could help to estimate the experimental bound on the values of parameters $\lambda$ and $K$.
	\end{abstract}
	 
	\section{Introduction} 
  		Nonlinear field theories find application in various fields of physics like quantum magnetism, the quantum Hall effect, meson theory, and string theory. One such example is non-linear sigma model. The introduction of the Skryme term in non-linear sigma models helps in the construction of static soliton solution in 3+1 dimensions \cite{1,1a,1b}. The exact solutions of nonlinear sigma or Skyrmion models are still difficult to find because of their highly non-linear character \cite{2}. However, using some ansatz one can considerably simplify such a system of equations to develop certain kind of solutions. One of such ansatz is hedgehog ansatz. The Einstein-Skyrme system which is a kind of self gravitating Skyrme model has been used by Droz, Heusler, and Straumann to obtain spherically symmetric black-hole solution numerically \cite{3}. Apart from these solutions with spherical symmetry, solutions with other symmetry have also been found \cite{4,5,6,7,8}. It is to be mentioned that all these solutions are found numerically. Thus no analytic solution was found there in Einstein-Skyrme system until Canfora and Maeda found one, using hedgehog ansatz \cite{9}. Some other interesting solutions constructed in context of Einstein-Skyrme theory are found in references \cite{sq,sw,se}.
  		
	 	The study of black hole thermodynamics started with the formulation of four laws black hole mechanics \cite{10}, and introduction of a black hole entropy \cite{11}. Hawking then found quantum evaporation of a black hole \cite{12,13,14}. In 1976 Gibbons and Hawking found that cosmological horizon can also radiate \cite{15}. As an important aspect of black hole physics, the study of black hole thermodynamics has played an important role in the development of quantum field theory in curved space-time. The study of black hole thermodynamics in curved space-time has its importance in understanding the very early universe \cite{16,17}.
	 	
	 	The correction to various thermodynamic quantities by quantum effects is an inevitable consequence of quantum gravity. Till now the thermodynamics of various black holes incorporating the quantum fluctuation have been studied in details. Some of the examples include the correction to Godel black \cite{18}. The study of quantum effects using Cardy formalism has also been discussed in \cite{19}. The detailed analysis of quantum corrections on thermodynamics of black holes has been discussed by Nozari et al \cite{20}. The effect of quantum corrections on thermodynamics of quasitopological black holes has bee done in \cite{21}. The study of quantum corrections to the thermodynamic of  Schwarzschild-Beltrami-de Sitter black hole \cite{22}, massive black hole in AdS space \cite{23},
	 	charged rotating AdS black holes \cite{23b}, charged black holes in gravity's rainbow
\cite{23c},	 	 dS black holes in massive gravity \cite{23a} have also been made so far. These corrections are necessarily logarithmic in nature \cite{24}. The detailed study in this regard of quantum corrections to black holes can be found in \cite{24a}. Some examples of further interest regarding quantum fluctuation have been carried in references \cite{24b,24c,24d,24e,24f,24g,24i,24j}.

		Motivated by recent work of constructing an analytic solution for the Einstein-Skyrmion system, we intend to discuss various thermodynamic variables of such a black hole in non-extended phase space. Firstly we compute ADM mass of Einstein-Skyrmion black hole using on shell Hamiltonian formalism. The thermodynamics of the black hole is then subjected to quantum fluctuations and we observe that the negative value of correction parameter $\alpha$ tends to increase the stability of the black hole system. The Helmholtz free energy shows minima for the negative value of the correction parameter $\alpha$. The blue curve in the pressure diagram \ref{pcplo} depicts that the negative value of correction parameter $\alpha$ induces dS-AdS phase transition in Einstein-Skyrmion black hole system. The Gibbs free energy also gets lowered for the negative value of $\alpha$. Then we discuss the $P-V$ criticality of the Einstein-Skyrmion black hole. From the plot for the temperature at different values of the horizon radius, we observe that there occurs a small to large black hole (SBH-LBH) transition, which one can see from figure 5. The $P-V$ criticality analysis done here reveals that despite obeying the criticality of a universal class of Van der Waals fluid, the ratio of critical constants does not turn out to be universal constant 3/8. The ratio thereof depends on the Skyrmion model parameters $\lambda$ and $K$. Hence we expect that our result can fix the bounds on the experimental values of $\lambda$ and $K$.

	The Plan of the paper is as following.  In section II, we would briefly discuss he Einstein-Skyrmion black hole solution and  and calculate its ADM mass. In section III, we discuss the the effects quantum fluctuations on various thermodynamic quantities of the Einstein-Skyrmion black hole. In section IV, we analyze the $P-V$ criticality of Einstein-Skyrmion black hole. In section V, the conclusion would be drawn.    
	\section{Einstein-Skyrmion black hole and its ADM mass}
	In this section, we give a short description of Einstein-Skyrmion black hole. The Skyrme Lagrangian  which describes the interactions of pions or baryons in low energy regime is now coupled to gravitational action. Now with the introduction of the SU(2) Skyrmion field, the Einstein-Skyrmion action becomes,
	\begin{equation}
	S = S_G + S_{Skyrmion},
	\end{equation} where $S_G$ gravitational action given by,
	\begin{equation}
	S_G= \frac{1}{16 \pi}\int d^4x \sqrt{-g}(R-2\Lambda),
	\end{equation} and the Skyrmion action is given by,
	\begin{equation}
	S_{Skyrmion} =\int d^4x \sqrt{-g} Tr\bigg (\frac{F_\pi^2}{16}R^\mu R_\mu + \frac{1}{32 e^2}F_{\mu \nu}F^{\mu \nu}\bigg ).
	\end{equation} Where $R_\mu$ and $F_{\mu \nu}$ are defined as,
	 \begin{eqnarray}
	R_\mu = U^{-1} \nabla U,\\
	F_{\mu \nu}=[R_\mu,R_\nu].
	\end{eqnarray} 
	The parameters $F_{\pi}$ and $e$ are fixed from experimental data. For convenience quantities $F_{\pi}$ and $e$ are  replaced by $K= F_{\pi}/4$ and $\lambda= 4/(e^2F_{\pi}^2)$ \cite{28}. With these changes the Skyrme action can now be written as,
	\begin{equation}
	S_{Skyrmion}=\frac{K}{2}\int d^4 x \sqrt{-g}Tr\bigg (\frac{1}{2}R^\mu R_\mu + \frac{\lambda}{16}F_{\mu \nu}F^{\mu \nu}\bigg ).
	\end{equation} One can now write the Einstein field equations as,
	\begin{equation}
	G_{\mu \nu}+\Lambda g_{\mu \nu}= 8 \pi T_{\mu \nu}.
	\end{equation}Here $G_{\mu \nu}$ is Einstein tensor and $T_{\mu \nu}$ is energy-momentum tensor. The parameterized form for Skyrmion action and energy-momentum tensor are given as,
	\begin{equation}
	\begin{aligned}
	S_{Skyrmion}= -K\int d^4x\sqrt{-g}\bigg [\frac{1}{2}G_{ij}(\nabla_\mu)Y^i(\nabla^\mu Y^j) \\ + \frac{\lambda}{4}[(G_{ij}(\nabla_\mu Y^i)(\nabla^\mu Y^j))^2 \\-G_{ij}(\nabla_\mu Y^i)(\nabla^\mu Y^j)G_{kl}(\nabla^\mu Y^k)(\nabla^\nu Y^l)]\bigg ]
	\end{aligned}
	\end{equation} and for energy momentum tensor we write,
	\begin{equation}
	T_{\mu \nu}= K \bigg [S{\mu \nu}-\frac{1}{2}g_{\mu \nu}S + \lambda[SS_{\mu \nu}-S_{\mu \alpha}S_{\nu}^{\alpha}-\frac{1}{4}g_{\mu \nu}(S^2-S_{\alpha \beta}S^{\alpha \beta})]\bigg ].
	\end{equation} Upto very recent for these field equations no analytic solution was present. However, many numerical solutions had already been there. Canfora and Maeda found an analytic solution \cite{9}, the metric for which was chosen to be spherically symmetric given by, 
	\begin{equation}
	ds^2=-f(r)dt^2+f(r)^{-1}dr^2+r^2(d\theta^2+\sin^2\theta d\phi^2),
	\end{equation} and the corresponding metric function is given by,
	\begin{equation}
	f(r)=1- 8 \pi K-\frac{2M}{r} +\frac{4 \pi K \lambda}{r^2} - \frac{1}{3} \Lambda r^2.
	\end{equation}
 	We now calculate the ADM mass of the Canfora and Maeda black hole solution. This particular solution has solid angle deficit due to Skyrmion matter parameters which is very similar deficits due to cosmic strings.  The Canfora and Maeda solution is quasi asymptotically flat with a deficit, so the integration parameter $M$ can be treated as Nucamendi– Sudarsky mass \cite{9a}. In reference \cite{9b}, using  quasi -local calculation \cite{9c,9d,9e,9f} the ADT mass for Skyrmion black hole  was calculated and is given as,
 	\begin{equation}
 	M_{ADT}= M(1-\alpha),
 	\end{equation} where $\alpha$ is the solid angle deficit parameter. To calculate ADM mass we use the formalism developed in \cite{9g}. We write the metric element of Skyrmion black hole in Schwarchild-AdS coordinates as, 
 	\begin{equation}
	 ds^2= - f(r)dt^2 +f(r) ^{-1}dr^2+C(r)d\theta^2+D(r)d\phi^2.
 	\end{equation} Here, 
 	\begin{eqnarray}
	f(r)=1+\frac{4 \pi  G \lambda  K}{r^2} -\frac{2 G M}{r}- \frac{1}{3} \Lambda r^2 &,&\notag \\
	C(r)= r^2(1- 8\pi G K)&,& \notag\\
	D(r)= r^2 \sin^2\theta(1- 8\pi G K).
 	\end{eqnarray} Using the formula for calculating mass of any black hole with unusual topology given in ref \cite{9g}, we write,
 	\begin{equation} \label{madm}
 	M_{ADM}= - \frac{1}{8 \pi}\lim\limits_{r \to \infty}\int_{S_g} \sqrt{f(r)}(\mathcal{H}-\mathcal{H}_0) \sqrt{\sigma}d\theta d\phi.\end{equation}
 	Here $\mathcal{H}$ is the trace of the extrinsic curvature of $S_g$ as embedded in a $t=const$ hyper-surface, $\mathcal{H}_0$ is the trace of extrinsic curvature of $S_g$ but now embedded in the reference space-time. Also $\sigma$ is two dimensional metric of the base manifold over which trace of $\mathcal{H}$ is computed. We first compute the trace as the covariant divergence of the normal vector field to the boundary. 
 	\begin{equation}\label{h}
 	\mathcal{H}= \sqrt{1 - \frac{1}{3} \Lambda r^2 - (\frac{2M}{r} -\frac{4 \pi G K \lambda}{r^2}) },
 	\end{equation} which in the limit of $r\to \infty$ can be binomially approximated to,
 	\begin{equation} \label{h0}
 	\mathcal{H}= \frac{2}{r}\sqrt{1-\frac{\Lambda r^2}{3}} + \frac{1}{r} \frac{(\frac{2 G M}{r}- \frac{4 \pi G K \lambda}{r^2})}{\sqrt{1-\frac{\Lambda r^2}{3}}}.
 	\end{equation} 
	 Similiarly we have,
 	\begin{equation}
 	\mathcal{H}_0= \frac{2}{r}\sqrt{1-\frac{\Lambda r^2}{3}} + \frac{1}{r} \frac{(\frac{2 G M_0}{r}- \frac{4 \pi G K 	\lambda}{r^2})}{\sqrt{1-\frac{\Lambda r^2}{3}}}.
	 \end{equation} Substituting \ref{h} and \ref{h0} in equation \ref{madm}, one can get the mass (for non unity genus $g$) as,
 	\begin{equation}
 	M_{ADM}= \frac{\pi(1- 8\pi G K) M }{4}.
 	\end{equation} Here $M$ is Nucamendi Sudarsky mass and we now calculate $M$ as the Nucamendi Sudarsky mass with the boundary located at event horizon and get the expression for ADM mass as,
 	\begin{equation}
  	 M_{ADM}= \frac{\pi(1- 8\pi G K)(12 \pi  G \lambda  K-24 \pi  K r^2-\Lambda  r^4+3 r^2)}{24 G r}.
 	\end{equation}The ADM mass formula we have obtained here is quite similar to the mass formula already derived for Einstein-Skyrme black hole using counter-term methods in Ref. \cite{9b}. There are some other black hole solutions in Einstein-Skyrme theory which have different underlying geometries. We can compare our results for ADM mass with  these black hole systems and can have better understanding of influence of geometry on black hole mass. For example, in Ref. \cite{se}, the mass formula for hairy black hole with flat horizon in Einstein $SU(2)$-Skyrme system has been derived. Also  the mass formula for the Einstein $SU(2)$-Skyrme system with black string (BTZ like)  is derived in \cite{se}.  Upon comparison with our result one can say that Einstein $SU(2)$-Skyrme system with black string is thermodynamically unstable. This thermodynamic instability is clear from the negative value of mass formula derived for Einstein $SU(2)$-Skyrme system with black string in Ref. \cite{se}.
	\section{Non extended Phase space thermodynamics}
	 We now briefly briefly discuss the thermodynamics of the Einstein-Skyrmion black hole. We would restrict ourselves to non extended phase space for which $\Lambda=0 $. We follow the formulation given in \cite{tt}. The Einstein-Skyrmion black hole solution in non extended phase space can be written as, 
	\begin{equation} \label{a}
		f(r)=1+\frac{4 \pi  G \lambda  K}{r^2}-8 \pi  G K-\frac{2 G M}{r}
	\end{equation} From here on we would for the sake of simplicity equate $G=1$. Firstly, we determine the event horizon by setting $f(r)$ in \ref{a} equal to zero. We have the roots as,
	\begin{equation}
	r_+=\frac{\sqrt{32 \pi ^2 \lambda  K^2-4 \pi  \lambda  K+M^2}+M}{1-8 \pi  K} 
		\hspace{.5cm}\text{and}\hspace{.5cm} r_-= \frac{M-\sqrt{32 \pi ^2 \lambda  K^2-4 \pi  \lambda  K+M^2}}{1-8 \pi  K}.
	\end{equation}
	Here $r=r_+$ denotes the outer horizon. 
	The temperature is calculated using, $$ T= \frac{f^\prime(r)}{4 \pi}\bigg|_{r=r_+}.$$ Thus for our case temperature becomes, 
	\begin{equation} \label{e}
	T=\frac{M r - 4 K \pi \lambda}{2 \pi r^3}.
	\end{equation}Here after without loss of generality we would use $r$ as the event horizon radius (i.e radius of outer horizon). All the calculations including $r$ are deemed for outer horizon.	
 	The mass term $M$ is given by,
 	 \begin{equation}
 	M=\frac{4 \pi  \lambda  K-8 \pi  K r^2+r^2}{2 r}.
 	\end{equation}The entropy of the black hole can be obtained using equation.
 	 \begin{equation}
 	S_0= \int \frac{dM}{T}= \int\frac{1}{T}\frac{\partial H}{\partial r}dr.
 	\end{equation}For the Einstein-Skyrmion black hole solution the entropy turns out to be,
 	\begin{equation}
 	S_0= \pi r^2.
 	\end{equation}
 	Thus in agreement with Bekenstein's area law. We now calculate the  Helmholtz free energy given by $F= - \int S_0 dT$. For Einstein-Skyrmion black hole the Helmholtz free energy is,
	\begin{equation} \label{fr}
	F=\frac{6 \pi  \lambda  K}{r}+M \log [r].
	\end{equation}Another thermodynamic quantity of interest is pressure. We derive expression for pressure of any thermodynamic system in non-extended phase space as $P= \frac{1}{2}TS$ \cite{tt}. Thus we have,
	\begin{equation}
	P=\frac{\left(4 \pi  \lambda  K-8 \pi  K r^2+r^2\right)-4 \pi  \lambda  K}{4 r}.
	\end{equation}By recognizing pressure and volume of black hole as conjugate to each other we can calculate the thermodynamic volume $V$ for the Einstein-Skyrmion black hole as, \begin{equation}
 	V= \frac{\partial H}{\partial P},
 	\end{equation} where $P$ is the pressure of the black hole system. Thus we get,
  	\begin{equation}
 	V= -\frac{4 \pi  r^2 \left(4 \pi  \lambda  K+(8 \pi  K-1) r^2\right)}{8 \pi  \lambda  K+(8 \pi  K-1) r^2}.
 	\end{equation} Where one can clearly see the manifestation of the Skyrmion model parameters $K$ and $\lambda$. We next derive the expression for internal energy as $U=H-PV=F+PV$ and we get,
 	\begin{equation}
	 U= \frac{4 \pi  \lambda  K+M r \log[r]}{r}.
	 \end{equation} Now we can calculate Gibbs free energy as $G= F+PV$ and we get the expression a,
	 \begin{equation}
	 G=\frac{\frac{\left(4 \pi  \lambda  K+(8 \pi  K-1) r^2\right) \left(8 \pi  \lambda  K+(8 \pi  K-1) r^2\right) (M r-6 \pi  \lambda  K)}{32 \pi ^2 \lambda ^2 K^2+(1-8 \pi  K)^2 r^4+16 \pi  \lambda  K (8 \pi  K-1) r^2}+12 \pi  \lambda  K+2 M r \log[r]}{2 r}.
 	\end{equation}Now we calculate the specific heat for Einstein-Skyrmion black hole, using the relation $ C= T (\frac{\partial S_0}{\partial T})$. Thus we get the expression for specific heat as,
 	\begin{equation}
 	C=-\frac{\pi  r^2 (M r-4 \pi  \lambda  K)}{M r-6 \pi  \lambda  K}.
 	\end{equation}
 	\section{Leading order quantum corrections to black hole thermodynamics} 
\begin{figure}[htb]
 	\begin{center}$
 		\begin{array}{c }
 		\includegraphics[width=80 mm]{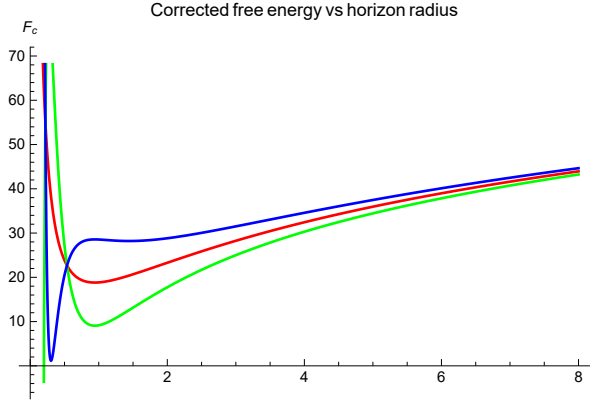}  
 		\end{array}$
 	\end{center}
 	\caption{ Corrected Helmholtz free energy w.r.t horizon radius for different values of correction parameter $\alpha$. Here red line corresponds to $\alpha =0$, green is for $\alpha=.5$ and blue is for $\alpha=-.5$.}
 	\label{f1}
 	 \end{figure}

\begin{figure}[htb]
  	\begin{center}$
  		\begin{array}{c }
  		\includegraphics[width=80 mm]{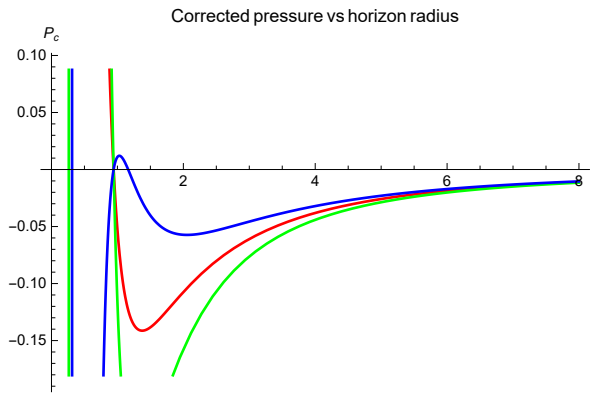}  
  		\end{array}$
  	\end{center}
  	\caption{ Corrected pressure w.r.t horizon radius for different values of correction parameter $\alpha$. Here red line corresponds to $\alpha =0$, green is for $\alpha=.5$ and blue is for $\alpha=-.5$.}
  	\label{pcplo}
  	\end{figure}

\begin{figure}[htb]
  	\begin{center}$
  		\begin{array}{c }
  		\includegraphics[width=80 mm]{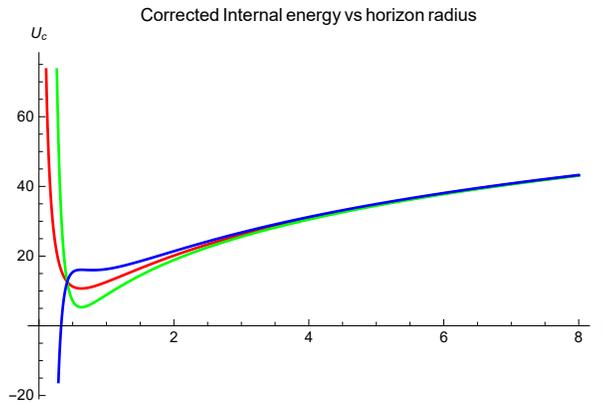}  
  		\end{array}$
  	\end{center}
  	\caption{ Corrected internal w.r.t horizon radius for different values of correction parameter $\alpha$. Here red line corresponds to $\alpha =0$, green is for $\alpha=.5$ and blue is for $\alpha=-.5$.}
  	\label{ucplo}
  \end{figure}

\begin{figure}[htb]
  	\begin{center}$
  		\begin{array}{c }
  		\includegraphics[width=80 mm]{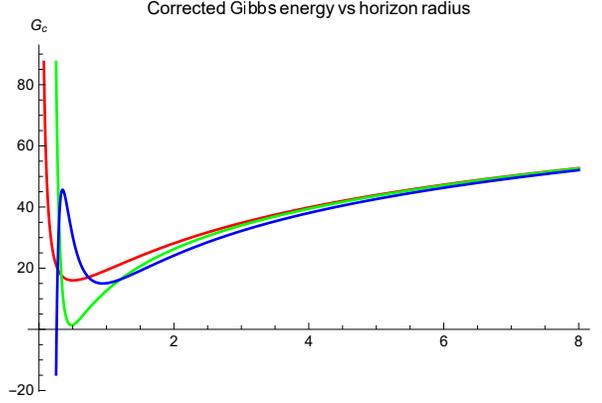}  
  		\end{array}$
  	\end{center}
  	\caption{ Corrected Gibbs energy w.r.t horizon radius for different values of correction parameter $\alpha$. Here red line corresponds to $\alpha =0$, green is for $\alpha=.5$ and blue is for $\alpha=-.5$.}
  	\label{gcplo}
\end{figure}

The entropy of black hole is changed by presence of small statistical thermal fluctuations around equilibrium thermodynamics. To observe this we first treat the black hole thermodynamic system as canonical ensemble and write down the expression for partition function. Trivially using definition of partition function we write,
 	\begin{equation}
	 Z(\beta) = \int_{0}^{\infty} dE \rho (E) e^{-\beta E},\end{equation} 
	 Using partition function we can write entropy as, \begin{equation}S(\beta) = lnZ + \beta E. 
	 \end{equation}We now expand $S(\beta) $ around the equilibrium value of temperature which is $\beta_0$, then the expansion becomes,
	 \begin{equation} \label{y}
 	S(\beta) = S_0 +\frac{1}{2}(\beta - \beta_0)^2 \frac{d^2S}{d\beta^2}|_{\beta = \beta_0} + \textnormal{higher order terms}.
 	\end{equation}We can generalize \ref{y} for different black hole geometries and write down general expression as,\begin{equation} \label{q}
	 S= S_0+ \alpha \log [S_0/4]+ \gamma/S_0+ \text{higher order terms}.
 	\end{equation}Here $S_0$ is the equilibrium or un corrected value of entropy. It should be noted that the correction parameters $\alpha$ and $\gamma$ enter as free parameters and depend on the black hole physics of particular black hole solution. It is found that $\alpha = −3/2 $ and $\gamma = −16/3$ for the BTZ black hole, $\alpha = −1 $ and $\gamma = −36/5$ for the AdS Schwarzschild black hole \cite{27}. In this paper we would consider only the first order corrections and ignore the higher order terms. To determine the qualitative effect of correction parameter on thermodynamics of Einstein-Skyrmion black hole, which is asymptotically AdS, we would take $\alpha=1/2$ \cite{la}.\\
 	We now use equation \ref{q} to obtain corrected thermodynamics due to first order quantum corrections. Thus the expression for entropy is now written as, 
 	\begin{equation}\label{w}
 	S=S_0+\alpha \log[4\pi r^2].
 	\end{equation}
 	 The corrected Helmholtz free energy ($F_c$) is obtained by using equations \ref{w} and \ref{e} in expression $F_C=\int S dT$ to obtain,
 	\begin{equation}
	 F_c=\frac{8 \pi  \alpha  \lambda  K-3 \alpha  \log \left[4 \pi  r^2\right] (M r-4 \pi  \lambda  K)+36 \pi ^2 \lambda  K r^2+6 \pi  M r^3 \log[r]-3 \alpha  M r}{6 \pi  r^3}.
	 \end{equation}
	 In the figure \ref{f1} we observe behavior of the Helmholtz free energy and obtain the effect
 	of the quantum corrections. We observe both positive and negative values of correction parameter $\alpha$. We observe that for negative value of $\alpha$ the minimum of free energy is well below the uncorrected curve as well as the green curve which corresponds to positive value of correction parameter $\alpha$. In the case of $\alpha = 0$ we recover the uncorrected Helmholtz free energy given in equation \ref{fr}. From figure we observe that the corrected and uncorrected curves cross at a particular point. Generally this point is related to the critical points. One should also observe that for large values of $r$ all the curves merge indicating that the quantum fluctuations tend to change the space-time structure only at lower values of horizon radius thus have their effect only for lower values of event horizon radius.\\
  As another important thermodynamic quantity we now discuss the effects of quantum fluctuation on pressure of Einstein-Skyrmion black hole. We denote the corrected pressure as $P_c$. The expression for $P_c$ is given by,
  \begin{equation}
  P_c =-\frac{\left[8 \pi  \lambda  K+(8 \pi  K-1) r^2\right]^2 \left[\alpha  \log \left[4 \pi  r^2\right]+\pi  r^2\right] (M r-6 \pi  \lambda  K)}{8 \pi ^2 r^5 \left[32 \pi ^2 \lambda ^2 K^2+(1-8 \pi  K)^2 r^4+16 \pi  \lambda  K (8 \pi  K-1) r^2\right]}
  \end{equation} In figure \ref{pcplo}, we observe the effects of quantum fluctuations on pressure of Einstein-Skyrmion black hole. We observe the the behavior of pressure curves for both corrected and uncorrected values from $r=1.2$ to $r=\infty$ is almost same. Also we can see that from $r=1.2$ up-to $\infty$, we see that the blue line for which correction parameter $\alpha$ is negative is well above the uncorrected red and corrected green curves. This indicates the quantum corrections have tendency to increase the cosmological constant. Hence, we see that quantum corrections can manifest in space time structure of the geometry. Also at $r=1.03$ the blue line tends to acquire positive value which is clear from the presence of blue peak above horizontal x-axis. This indicates that there is a transition of black hole from AdS to dS space-time. The order of this phase transition would be clear when we will discuss specific heat of Einstein-Skyrmion black hole. The original space time structure and cosmological constant of Einstein-Skyrmion black hole is seen from the red curve. One can also see that all the three curves which are red, green and blue merge below $r=.3$. This indicates that space-time structure is completely opaque to any corrections at very small values of horizon radius. One can say that the formation and transition of black holes from large size to small sized black holes (or vice-versa) seizes here. \\
  We now discuss the effects that the quantum corrections tend to bring in the internal energy of Einstein-Skrymion black hole. We denote the corrected internal energy by $U_c$.
  \begin{equation}
  U_c = \frac{8 \pi  \lambda  K \left(\alpha +3 \pi  r^2\right)-3 \alpha  M r}{6 \pi  r^3}+M \log [r]
  \end{equation} In figure \ref{ucplo} we see that from $r = \infty$ up to a value of $r=0.3$ blue line has tendency to increase the internal energy. This is the indication of stability. However internal energy shows an asymptotic decrease at value of $r=0.3$. This effect we attribute to the AdS to dS transition, which happens at this particular value of horizon radius as seen earlier in the pressure curve. The behavior of internal energy for positive value of $\alpha$ and $\alpha=0$ is expectedly same. \\
 We now discuss the effects of quantum fluctuations on Gibbs free energy of Einstein-Skrymion black hole. The corrected Gibbs free energy denoted by $G_c$, is given as,
 \begin{eqnarray}
 G_c=\frac{\frac{3 \left(4 \pi  \lambda  K+(8 \pi  K-1) r^2\right) \left(8 \pi  \lambda  K+(8 \pi  K-1) r^2\right) \left(\alpha  \log \left[4 \pi  r^2\right]+\pi  r^2\right) (M r-6 \pi  \lambda  K)}{32 \pi ^2 \lambda ^2 K^2+(1-8 \pi  K)^2 r^4+16 \pi  \lambda  K (8 \pi  K-1) r^2}}{6 \pi  r^3}  \notag\\
 +\frac{8 \pi  \alpha  \lambda  K-3 \alpha  \log \left[4 \pi  r^2\right] (M r-4 \pi  \lambda  K)+36 \pi ^2 \lambda  K r^2+6 \pi  M r^3 \log [r]-3 \alpha  M r}{6 \pi  r^3}.
 \end{eqnarray} The effects of quantum fluctuations on Gibbs free energy are depicted in figure \ref{gcplo}. From the begining we have observed that the negative value of quantum correction parameter $\alpha$ leads to the stability. We expect $\alpha$ to decrease the Gibbs free energy as well. The blue curve in figure 5 has tendency to decrease the Gibbs free energy. We also see a blue spike at a value of $r=0.3$. We attribute this to the AdS to dS transition.    Next we discuss the specific heat which for any thermodynamic system is very important quantity that gives idea about the stability and viability of the system. Also the specific  heat of any thermodynamic system  has a say about order of the  phase transition (if any)  happening in that system. We derive the expression for corrected specific heat of Einstein-Skyrmion black hole by using equations  \ref{e}and \ref{w}. The corrected specific heat, denoted by $C_c$ is the given by\begin{equation} \label{r}
C_c=-\frac{\left(\alpha +\pi  r^2\right) (4 \pi  \lambda  K-M r)}{6 \pi  \lambda  K-M r}.
\end{equation}The specific heat of Einstein-Skyrmion black hole shows only one discontinuity, see the red curve in figure 5. Also the blue and green curves too have only one discontinuity each indicating that there is only first order phase transition in Einstein-Skyrmion black hole. Also it is to be noted that the curve of specific heat drawn here is only a qualitative indications. We have kept the model parameters $K$ and $\lambda$ silent by equating them to 1. To identify the regions of stability on must solve the equation \ref{r} for $C_c >0$.
  \begin{figure}[htb]
	\begin{center}$
		\begin{array}{c }
		\includegraphics[width=80 mm]{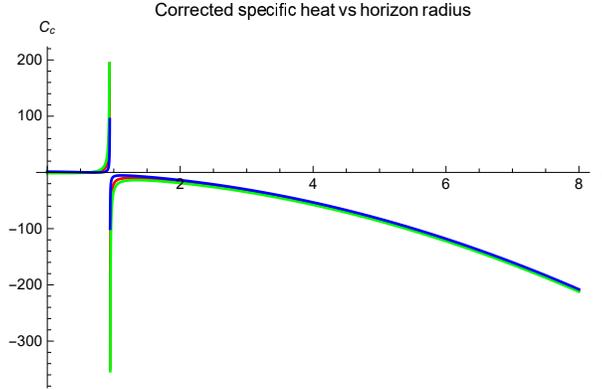}  
		\end{array}$
	\end{center}
	\caption{ Corrected specific heat w.r.t horizon radius for different values of correction parameter $\alpha$. Here red line corresponds to $\alpha =0$, green is for $\alpha=.5$ and blue is for $\alpha=-.5$.}
	\label{shplo}
	\end{figure}
	\section{$P-V$ Criticality of Einstein-Skyrmion black hole}
	By treating cosmological constant as pressure of a black hole system, one can discuss some nice features in black hole thermodynamics such as phase transition, $P-V$ criticality, heat cycles etc. Including cosmological constant in metric function one can write the expression for metric function of Einstein-Skyrmion black hole as 
	\begin{equation}
	f(r)=1+ \frac{4 \pi  \lambda  K}{r^2}-8 \pi  K-\frac{2 M}{r}-\frac{\Lambda  r^2}{3}.
	\end{equation}Now taking $\Lambda= \frac{-3}{l^2},$ where $l$ is AdS length parameter, we derive the expression for temperature as,  
	\begin{equation} \label{i}
	T=-\frac{8 \pi  \lambda  K}{r^3}+\frac{2 r}{l^2}+\frac{2 M}{r^2}.
	\end{equation}Choosing various parameters appropriately we plot temperature for different values of horizon radius. We see that for different values of horizon radius there exist different black holes, the temperature of these different black holes have different behavior. This indicates that there there exists a small-large
	black hole (SBH-LBH) transition in the extended phase space. Such a transition for different black hole solutions like Reissner-Nordstr¨om black hole is discussed in \cite{25,26}. 
		\begin{figure}[htb]
		\begin{center}$
			\begin{array}{c }
			\includegraphics[width=80 mm]{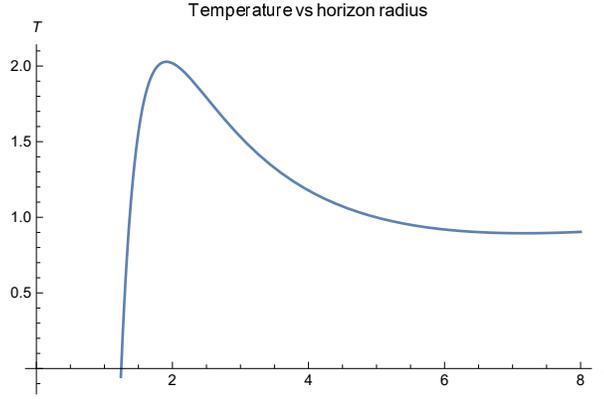}  
			\end{array}$
		\end{center}
		\caption{ Variation  of temperature $T$ w.r.t $r$ }
		\label{TVR}
	\end{figure}  
	 We discuss $PV$ criticality Einstein-Skrymion black hole by starting  with equation of state as: 
	\begin{equation} \label{o}
P=\frac{3 \pi  \lambda  K}{2 r^4}+\frac{3 \pi  k}{r^2}-\frac{3}{8 \pi  r^2}+\frac{3 T}{2 r}.
	\end{equation}
	 We determine critical points using the condition,
	 \begin{equation}
	\frac{\partial P}{\partial r}=0, \hspace{1cm} \frac{\partial^2 P}{\partial r^2}=0, 
	 \end{equation} which leads to, 
	 \begin{equation} \label{p}
	 \begin{aligned}
	 T_{critical}= \frac{\left(1-8 \pi ^2 K\right)^{3/2}}{6 \sqrt{6} \pi ^2 \sqrt{K} \sqrt{\lambda }},\hspace{3cm} r_{critical} =\frac{2 \sqrt{6} \pi  \sqrt{K} \sqrt{\lambda }}{\sqrt{1-8 \pi ^2 K}},\\
	 P_{critical}=\frac{\left(1-8 \pi ^2 k\right)^2}{128 \pi ^3 K \lambda }\hspace{1cm} V_{critical} =\frac{64 \sqrt{6} \pi ^4 K^{3/2} \lambda ^{3/2}}{\left(1-8 \pi ^2 K\right)^{3/2}}.
	  \end{aligned}
	 \end{equation}Thus,
	 \begin{equation}
	 \frac{P_{critical}V_{critical}}{T_{critical}}= \frac{18 \pi ^3 K \lambda }{1-8 \pi ^2 K}.
	 \end{equation}
	 We know from \cite{9b} that the critical behavior of Einstein-Skyrmion black hole is in accordance with the critical behavior of a Van der Waals fluid. Thus we can equate the above ratio to a universal constant 3/8. We therefore can estimate the bound for the experimental values of $K$ and $\lambda$ using the equation,
	 \begin{equation}\label{t}
	 \frac{18 \pi ^3 K \lambda }{1-8 \pi ^2 K}=\frac{3}{8}.
	 \end{equation}One can use equation \ref{t} to determine the values of the parameters $\lambda$ and $K$. These constants appear in theory when one deals with the strong interactions of baryons and mesons in Quantum Chromodynamics (QCD) \cite{90}.
	 \section{Conclusion}
	 In this paper, we discussed the thermodynamics of a recently constructed analytic solution in the Einstein-Skyrmion theory. The leading order corrections to the thermodynamics of Einstein-Skyrmion were carried in section III. We retained our analysis to the first order. The corrections we discussed are quantum fluctuations that arise in the quantum gravity regime. We observe that the negative value of the correction parameter $\alpha$ has a tendency to increase the stability of the black hole system. This is clear from the entropy equation \ref{e}. First, we analyze the effects of quantum fluctuations on the Helmholtz free energy and plot them in figure \ref{f1}. The blue curve which corresponds to the negative value of correction parameter $\alpha$ has the minimum well below the other curves indicating the stability. Next, we draw in figure \ref{pcplo}, the effects of quantum corrections on the pressure of Einstein-Skyrmion black hole. The pressure plot indicates that for the negative value of correction parameter $\alpha$, we see a dS-AdS phase transition, see the blue curve. The internal energy subjected to quantum correction is drawn in figure \ref{ucplo}. The Gibbs free energy and the effects of quantum corrections on Gibbs free energy of the Einstein-Skyrmion black hole are given in figure \ref{gcplo}. Apart from the region where we see dS-AdS phase transition, the negative value of correction parameter $\alpha$ has always a tendency to lower the Gibbs free energy of the system, indicating the stability. Another important thermodynamic quantity which is specific heat is given in figure \ref{shplo}. We see that the red, green, and blue curves which correspond to three different values of correction parameter $\alpha$, have one discontinuity each. This indicates the there is only the first-order phase transition. The dS-AdS phase transition which was seen in the pressure curve is thus a first-order phase transition. In section IV we discuss the $P-V$ criticality of the Einstein-Skyrmion black hole. We plot the temperature for different values of the horizon radius. We observe that there occurs a small to large black hole (SBH-LBH) transition, which is clear from the figure \ref{TVR}. We then include the cosmological constant in metric function and obtain the expression for temperature given in equation \ref{i}. By recognizing the cosmological constant as the pressure of the black hole system, we then derive the equation of state for Einstein-Skyrmion black hole, given in \ref{o}. From here we derive the critical constants and denote critical temperature, radius,  pressure, and volume by $T_{critical}$, $r_{critical}$, $P_{critical}$ and $V_{critical}$ respectively. Equation \ref{p} gives the expressions for  $T_{critical}$, $r_{critical}$, $P_{critical}$ and $V_{critical}$. Next we determine the ratio $\frac{P_{critical} V_{critical}}{T_{critical}}$. We see that this ratio does not turn out to be a constant number but depends on the Skyrmion model parameters $\lambda$ and $K$. We know the $P-V$ criticality class of Einstein-Skyrmion black holes is Van der Waals. Thus equating the ratio of critical constants given in equation \ref{a}, we can have bound on the experimental values of the model parameters $\lambda$ and $K$.
	 	
\end{document}